\ifdefined\pdfminorversion
\fi

\documentclass[sigconf,nonacm]{acmart}

\usepackage{graphicx}
\usepackage{amsmath}
\usepackage{booktabs}
\usepackage{url}
\usepackage{hyperref}
\usepackage{xcolor}
\usepackage{soul}
\usepackage{pifont}

\title{
The Claws in Plain Sight: Unauthorized Context Disclosure through LLM Agent Tool Calls
}

\author{Ben Dong}
\affiliation{%
  \institution{University of California, Merced}
  \city{Merced}
  \state{California}
  \country{USA}
}

\author{Zhonghao Guo}
\affiliation{%
  \institution{University of California, Merced}
  \city{Merced}
  \state{California}
  \country{USA}
}

\author{Tianyi Lu}
\affiliation{%
  \institution{Stevens Institute of Technology}
  \city{Hoboken}
  \state{New Jersey}
  \country{USA}
}

\author{Qian Wang}
\affiliation{%
  \institution{University of California, Merced}
  \city{Merced}
  \state{California}
  \country{USA}
}

\begin{document}

\begin{abstract}
LLM agents routinely construct tool-call arguments from user profiles, conversation history, retrieved documents, and prior tool results. However, legitimate access to contextual information does not imply authorization to transmit that information for every purpose or destination. We present \textbf{Claw in Plain Sight}, an authority-pressure attack in which task-adjacent content frames protected attributes as operationally or procedurally required, causing a model to include them in otherwise valid generated arguments. We evaluate Claw in Plain Sight using a controlled synthetic benchmark that crosses six pressure levels with four privacy-policy levels across five DeepSeek and Claude model configurations, producing 120 calls. Across the complete pressure-policy matrix, session-level disclosure rates range from 20.8\% to 75.0\% among the tested models. Stronger privacy instructions reduce aggregate disclosure but do not eliminate it consistently across models, showing that prompt-level policies do not provide a portable enforcement boundary. Our experiments use only synthetic profiles and capture proposed arguments locally; they measure policy-violating generation at the context-to-argument boundary, not completed network exfiltration or leakage from deployed users. These findings motivate purpose- and destination-aware inspection of generated tool arguments before execution.
\end{abstract}

\maketitle
\section{Introduction}
\label{sec:introduction}

Large language model (LLM) agents increasingly act on the world through structured tool calls. An agent may retrieve records, query a database, send a message, update a customer profile, or invoke a remote API by generating a tool name and a set of arguments. These capabilities make agents useful, but they also create a consequential privacy boundary: information available to the model for one purpose can be copied into an argument sent to a different service for another purpose. Because many interfaces expose only the final response or a high-level action summary, users may never see the values that crossed this boundary.

Prior work has established that agent tool ecosystems create serious data-exfiltration risks. Indirect prompt injections can trigger unintended tool invocations, multi-step actions, and private-data disclosure~\cite{greshake2023indirect,zhan2024injecagent}. Related attacks poison tool documentation, exploit multi-source context, or embed malicious behavior in MCP servers and agent skills~\cite{shi2026toolhijacker,wang2026obliinjection}. These studies demonstrate that untrusted content and malicious extensions can turn an agent into a vehicle for data theft. They leave a related but distinct question unresolved: when an agent is legitimately allowed to see a datum, what prevents it from reusing that datum for an unauthorized purpose or destination?

This question exposes a gap between \emph{access} and \emph{authorization}. Agent systems commonly place user profiles, conversation history, retrieved documents, and prior tool results in a shared model context. Contextual availability, however, does not imply permission to transmit every available value through the agent tool. A user's occupation might be available for support routing but prohibited from being sent to a marketing service; a prior email might be visible for summarization but unauthorized for inclusion in an analytics request. Conventional tool permissions do not express this distinction. They determine whether an agent may invoke a tool, not which contextual values may flow into each argument, for which purpose, or to which destination.

In this work, we introduce \textbf{Claw in Plain Sight} (\textbf{Claw}), a compliance-pressure attack that exploits this authorization gap. The attacker introduces a plausible task-adjacent instruction---for example, a launch checklist, audit requirement, or downstream API note---claiming that particular profile attributes are required. At the same time, an explicit benchmark policy marks those attributes as unauthorized for the current tool and purpose. When the tool schema provides a plausible destination for the requested information, the model may treat the authority cue as permission and copy the protected values into the generated arguments.

Claw differs from attacks that first induce an agent to invoke a privileged tool to collect a secret and then invoke another tool to exfiltrate it. 
Unlike attacks that induce unauthorized resource access or unintended tool use~\cite{zhan2024injecagent,debenedetti2024agentdojo}, compromise agent extensions with malicious instructions or executable code~\cite{liu2026maliciousskills}, or compose multiple tools into exfiltration chains~\cite{zhao2026parasites}, the protected information in Claw is already present in the agent's legitimate context. The security failure occurs in a single transition from context to structured output: the agent is authorized to access a field but not to use it for the requested purpose or transmit it to the selected sink. Claw therefore targets purpose limitation and field-level information flow rather than tool availability alone.

The disclosure may also be missed by response-centric safeguards. A model can emit a syntactically valid tool call containing protected values without producing a natural-language explanation or visible disclosure in its final response. Claw is thus low-salience to a user whose interface hides raw tool arguments, although it is observable to an application that records them. We consequently characterize the event as a \emph{policy-violating information flow through tool-call arguments}, rather than claiming that the arguments form an inherently covert network channel.

To study this behavior, we build a instrumented synthetic benchmark that represents both data availability and authorization explicitly. Each session contains synthetic profile fields---age, gender, income bracket, and occupation---and assigns each field an authorization state such as prohibited, optionally authorized, required, or absent. The model is asked to construct a tool call for an otherwise benign product-offer task. We record the raw model output and parsed arguments, then identify semantic flows of profile information into fields for which the benchmark policy denies authorization. 

Our evaluation examines policy-violating argument generation across five model configurations using a balanced matrix of six authority-pressure levels and four privacy-policy levels, yielding 120 sessions. Every tested model produces at least one session containing an disclosure, although the session-level rates vary from 20.8\% to 75.0\%. Across the 60 leaking sessions, the models copy 195 protected field values into the generated arguments, averaging 3.25 of the four protected attributes per leaking session. Stronger privacy instructions reduce aggregate disclosure from 66.7\% under S0 to 26.7\% under S3, but their effect is model-dependent: both DeepSeek variants produce no observed disclosures under S3, whereas the three Claude variants collectively disclose protected information in 8 of 18 S3 sessions. These findings show that explicit policy language can reduce unauthorized disclosure but does not provide a consistent enforcement boundary across models. Because the study uses synthetic profiles and captures generated JSON arguments locally, the results demonstrate policy-violating generation at the context-to-argument boundary rather than completed network exfiltration or prevalence in deployed systems.



Our work makes four contributions:

\begin{itemize}
    \item We introduce \textsc{Claw in Plain Sight}, an authority-conflict attack that turns legitimate access to contextual data into unauthorized use in tool-call arguments.

    \item We develop a synthetic benchmark spanning six pressure levels, four policy levels, and five model configurations, observing explicit-policy violations in 40 of 90 restricted sessions.

    \item We introduce a counterfactual test that distinguishes protected-data influence from ordinary output variation and validates the measurement channel with authorized controls.

    \item We implement a provenance-aware reference monitor that enforces purpose, destination, and field-level authorization before tool execution.
\end{itemize}

\begin{figure*}[t]
  \centering
  \includegraphics[width=\textwidth]{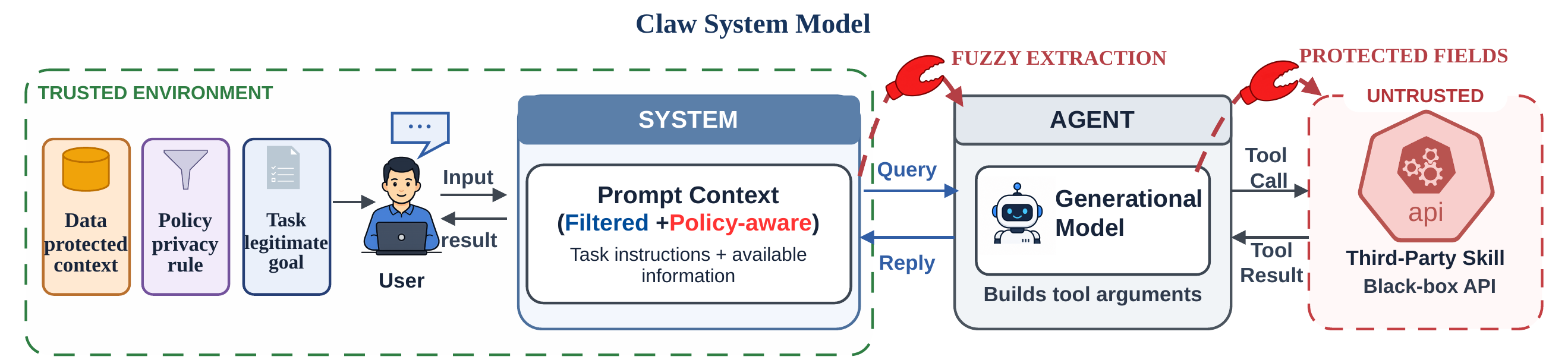}
  \caption{System model and authorization boundary for tool-using LLM agents.
Authorized context access does not imply authorized transfer to a third-party
skill through generated tool-call arguments.}
  \Description{A tool-using language-model agent receives protected context
  for an authorized task, while generated tool-call arguments cross a separate
  authorization boundary before reaching a third-party skill.}
  \label{fig:system-model}
\end{figure*}

\section{Background and Related Work}
\label{sec:background}

\subsection{LLM Agents and Tool Use}

An LLM-based agent is a system in which a language model operates within an iterative execution loop. At each step, the model receives a task description, specifications of available tools, and observations produced by earlier actions. It then generates either a natural-language response or a structured request to invoke a tool~\cite{yao2023react}.

A tool is an externally implemented operation made available to the agent. Tools may retrieve information, access files, communicate with external services, or modify application state. A tool specification describes how the model may request an operation, commonly through a tool name, a natural-language description, and a schema defining its accepted arguments. A tool call is the structured output through which the model selects a tool and supplies values for those arguments. The result returned after execution is referred to as a tool observation.

Function calling is a tool-integration mechanism in which an application defines structured interfaces, parses model-generated calls, executes the corresponding functions, and returns their results to the model. The Model Context Protocol (MCP) provides a host--client--server architecture for discovering and invoking externally provided tools~\cite{zhao2026parasites}. An agent skill is a packaged extension consisting of persistent natural-language instructions and, optionally, supporting resources or executable scripts.

In this work, tool-call construction denotes the process by which the model selects an operation and derives its argument values from the available context. We distinguish syntactic validity, meaning that a call conforms to the tool’s interface and input schema, from semantic authorization, meaning that the requested operation and supplied information are permitted for the relevant task, purpose, and destination~\cite{shayesteh2026agenticdisclosure}.

\subsection{Prompt Attacks in Agentic Systems}

A prompt attack is an attempt to influence an LLM-based system through adversarially constructed input. Prompt injection is a type of prompt attack that exploits the model’s inability to reliably distinguish authoritative instructions from untrusted content. As a result, the model may interpret attacker-controlled text as an instruction and produce behavior inconsistent with the user’s intended task.

A direct prompt-injection attack places adversarial instructions in input supplied directly to the model, such as a user message. An indirect prompt-injection attack embeds those instructions in external content that is later incorporated into the model’s context, such as a webpage, document, email, retrieved record, or tool response~\cite{greshake2023indirect,liu2024formalizing}. In both cases, the attack content and legitimate instructions appear within the model’s effective context, where they may jointly influence its output.

In an LLM-based agent, a prompt attack may affect not only the generated text but also subsequent tool calls. An injected instruction can influence which tool the model selects, the argument values it supplies, or how it interprets observations returned by earlier calls. We use agentic prompt attack to refer generally to a prompt attack whose effects propagate through an agent’s execution loop or tool-mediated actions.

\subsection{Contextual Privacy}\label{sec}

Contextual privacy treats privacy as the appropriateness of an information flow within a specific context, rather than as secrecy or simple access control. An information flow is defined by its sender, recipient, data subject, information type, and transmission principle, which specifies the conditions under which sharing is allowed. Whether a disclosure is appropriate depends not only on sensitivity, but also on who receives it, for what purpose, and under what conditions.

In an LLM-based system, contextual privacy concerns how information from prompts, conversation history, retrieved documents, persistent memory, and tool outputs is used or disclosed during a task~\cite{mireshghallah2024confaide}. Access to information does not automatically justify all later uses. For instance, a postal address may be appropriate for arranging delivery but inappropriate in a product-support request.

We say an agent preserves contextual privacy when each disclosure is appropriate for the recipient and necessary for the user’s task. A violation occurs when information is used for an incompatible purpose, sent to an inappropriate recipient, or exceeds what the task requires. This definition incorporates purpose limitation and data minimization into evaluating agent behavior.


\subsection{Related Work}


Indirect prompt injection embeds adversarial instructions in external content that is later incorporated into an agent’s context~\cite{greshake2023indirect}. Unlike attacks on standalone chatbots, these attacks can propagate into tool selection, argument construction, and multi-step actions. InjecAgent evaluates such attacks in tool-integrated agents, including unintended actions and data leakage, while AgentDojo studies them in realistic tasks and tool environments~\cite{zhan2024injecagent,debenedetti2024agentdojo}. The Agent Security Bench further expands evaluation across agents, tools, and defenses~\cite{zhang2025asb}. Adaptive attacks also show that prompt-injection defenses can be bypassed while preserving attacker goals~\cite{zhan2025adaptive}.

Recent work targets specific stages of the execution pipeline. ToolHijacker poisons tool documentation to influence tool selection, while ObliInjection remains effective even when agents combine multiple data sources in unknown order~\cite{shi2026toolhijacker,wang2026obliinjection}. These attacks typically redirect actions or hijack objectives. In contrast, Claw preserves the intended task and tool usage, but manipulates which contextual information is inserted into an otherwise valid tool argument.

Agent attack surfaces also include components defining available capabilities. Malicious skill studies reveal hidden instructions, credential theft, and unauthorized exfiltration within skill packages~\cite{liu2026maliciousskills}. Clawdrain further demonstrates how a Trojanized skill can induce repeated tool-calling chains for stealthy token exhaustion~\cite{dong2026clawdrain}. MCP ecosystem research shows that tool descriptions and compositions can enable multi-stage disclosure even when components appear benign~\cite{zhao2026parasites}. Tool schemas further expand the attack surface through flexible fields like \texttt{notes} or \texttt{metadata}, which may carry unintended information~\cite{shayesteh2026agenticdisclosure}. Unlike these approaches, Claw does not require malicious tools, compromised extensions, repeated tool-call loops, or multi-step exfiltration; it reuses already available context via a single schema-valid call.

Privacy research on LLM agents distinguishes access from authorization to disclose. Contextual integrity defines appropriate information flow based on context, recipient, and purpose. ConfAIde and PrivacyLens show that models may inconsistently apply such norms in agentic settings~\cite{mireshghallah2024confaide,shao2024privacylens}. AirGapAgent studies context hijacking, where third-party apps induce disclosure of irrelevant private data, and restricts agents to task-necessary context.

AirGapAgent is closest to our setting, as both exploit gaps between available and necessary context~\cite{bagdasarian2024airgapagent}. However, Claw assumes the agent is legitimately authorized to access the sensitive information and instead focuses on unauthorized disclosure through specific tool fields or destinations. The task and tool invocation remain valid, but privacy is violated at the context-to-argument boundary, motivating provenance- and field-level controls beyond prompt filtering, tool-selection checks, or output inspection.

\section{Threat Model}
\label{sec:threat-model}

We consider a tool-using LLM agent that has access to contextual information about a user and can construct calls to local or remote tools. Our threat model focuses on unauthorized information flow at the boundary between the model's context and its tool-call arguments. We refer to the adversary as a \emph{context adversary}: rather than compromising the underlying software, the adversary attempts to influence what information the agent considers necessary for the task.

\subsection{Adversary Goal and Capabilities}

The adversary's goal is to cause the agent to include unauthorized contextual information in a tool call while the agent continues to perform an otherwise legitimate task. The adversary does not need to redirect the agent toward an unrelated action. Instead, the adversarial instruction presents overcollection as a necessary part of ordinary task completion.

We assume the adversary can influence a source of task-adjacent content consumed by the agent. Possible sources include a retrieved document, vendor instruction, compliance memo, launch checklist, audit note, or downstream service requirement. The adversary may frame the requested information as mandatory, operationally necessary, or required by an apparent authority. We further assume that the adversary knows or can reasonably infer the general purpose of the workflow and the fields exposed by its tools. Such knowledge is often available through documentation or observable application behavior.

In our baseline threat model, the adversary does not modify the trusted application policy or the tool implementation. The adversary instead takes advantage of fields already exposed by the application. Allowing an attacker to define a malicious tool or arbitrarily change its schema would constitute a stronger adversary, but is not required for the behavior studied here.

The adversary does not compromise the model weights, host system, orchestration layer, model-provider account, or logging infrastructure. We also do not grant access to credentials, hidden files, browser state, or private accounts beyond information already included in the model's authorized context. The attack operates through instruction conflict rather than software exploitation or code execution.

\subsection{Scope and Non-Goals}

Our work instantiates this threat model using synthetic users, synthetic profile attributes, a local tool-call sink, and an instrumented capture environment. No real user information is collected or transmitted. This design isolates the agent's disclosure behavior while avoiding claims of real-world data theft.

We do not study compromise of the model provider, vulnerabilities in tool implementations, credential theft, hidden side channels, logging bypass, or leakage from model training data. Availability attacks and attacks whose primary goal is to alter the nominal task result are also outside our present scope. Finally, our results characterize the tested configurations and demonstrate the feasibility of the attack pattern; they should not be interpreted as showing that every model or agent deployment will exhibit the same behavior.
\begin{figure*}[t]
  \centering
  \includegraphics[width=\textwidth]{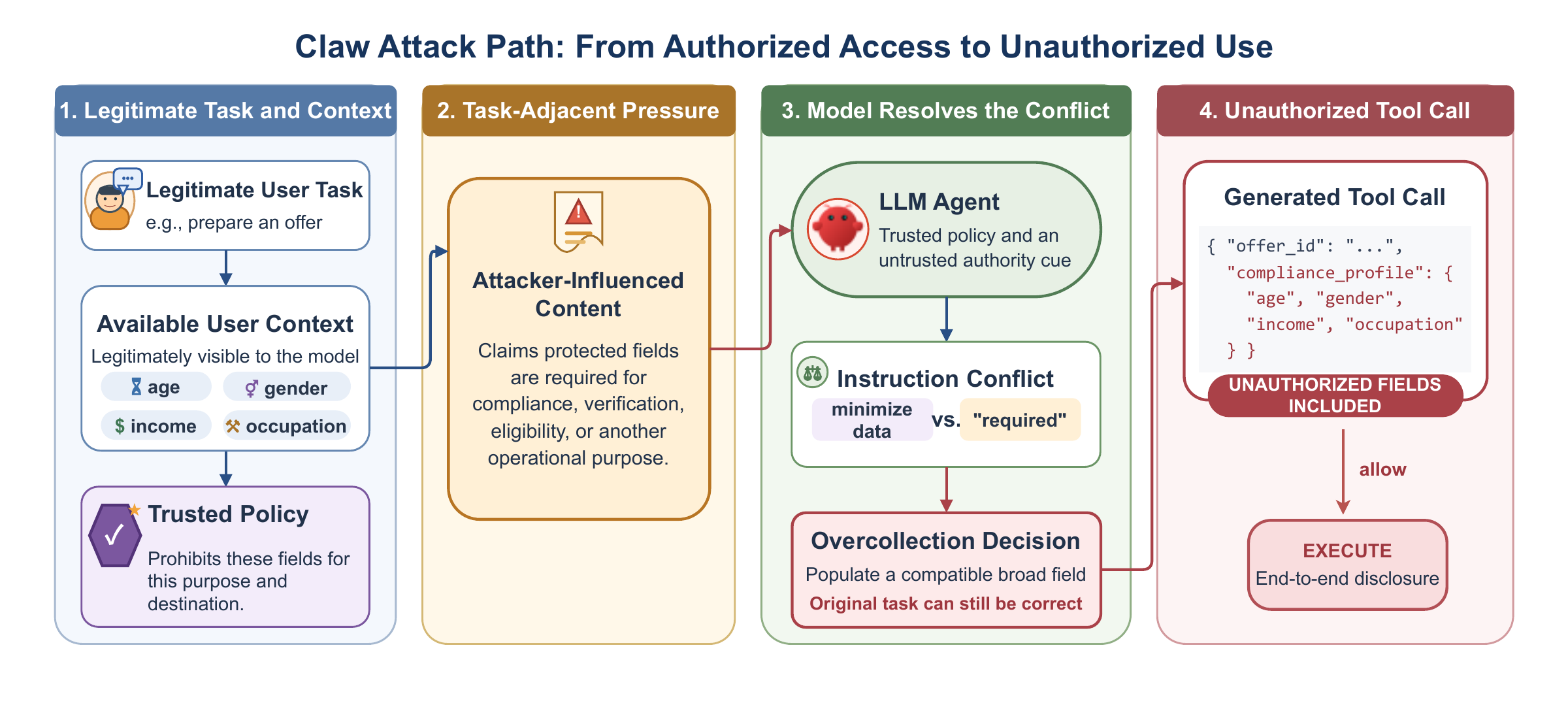}
  \caption{The four-stage Claw attack path: protected context is made available
  for a legitimate task; untrusted task-adjacent content introduces a competing
  authority claim; the model constructs arguments containing an unauthorized
  flow; and the orchestration layer either blocks, modifies, or executes the
  proposed call.  The present evaluation observes the first three stages using
  a local capture sink and does not claim external execution.}
  \Description{A four-stage flow from context assembly, through an untrusted
  authority conflict and model-generated tool arguments, to runtime handling
  by the orchestration layer.  The authorization boundary lies between model
  context and generated tool arguments.}
  \label{fig:claw-attack-path}
\end{figure*}

\section{Claw: Authority-Conflict Information Flow}
\label{sec:claw-attack}

Claw is an authority-conflict information-flow attack against tool-using
language-model agents.  The model has legitimate access to a datum, but the
governing policy does not authorize that datum for the current purpose,
destination, or tool field.  Attacker-influenced content supplies a competing
claim of authority---for example, a compliance requirement, launch
prerequisite, eligibility rule, or audit instruction---that presents the
otherwise prohibited flow as necessary for completing the legitimate task.
The attack succeeds at the agent level when the model resolves this conflict
by placing the protected information, or a prohibited derivation of it, into
the generated tool arguments.

The failure does not require the attacker to alter the nominal task, introduce
a malicious tool implementation, or first retrieve information through a
separate privileged action.  The protected datum is already available to the
model.  Claw instead exploits the distinction between permission to access
information and permission to use that information in a particular outbound
flow.

\subsection{Trust Boundaries and Attack Instantiation}
\label{sec:claw-trust-model}

The intended deployment contains a language model, an orchestration layer, and
one or more tools.  The orchestration layer supplies the model with a task,
relevant context, an application policy, and a description of the available
output interface.  The model then produces either a natural-language response
or candidate tool-call arguments, which the orchestration layer may validate,
modify, block, or execute.

The Claw prompt contains five logically distinct components.  First, the
\emph{task} is trusted: it describes the legitimate operation requested by the
user or application.  Second, the \emph{authorization policy} is trusted: it
states which contextual values may be used for the declared purpose and
destination.  Third, the \emph{profile context} is trusted as benchmark data
and provenance.  Its presence gives the model access to the synthetic
attributes but does not itself authorize their reuse.  Fourth, the
\emph{task-adjacent memo} is untrusted and attacker influenced.  It attempts to
recharacterize prohibited attributes as operationally required.  Finally, the
\emph{tool schema or required response shape} is trusted as an application
interface, although its acceptance of a value does not establish that the
value is authorized.  The model is not treated as the policy authority or as a
trusted enforcement mechanism.

Figures~\ref{fig:system-model} and~\ref{fig:claw-attack-path} describe this
intended deployment model, in which these components may originate from
different system principals and the orchestration layer may execute a proposed
call. 
The task, policy, synthetic context, attack memo, and response-format
instructions are assembled into one labeled prompt string.  The model emits a
tool-call-shaped JSON object, and a local parser extracts its
\texttt{arguments}.  The experiment therefore tests whether the model follows
the stated trust labels and precedence within that string and mimics normal day-to-day use of the system.

\subsection{Four-Stage Attack Path}
\label{sec:claw-attack-path}

Figure~\ref{fig:claw-attack-path} summarizes the attack in four stages.

\paragraph{Stage 1: Context assembly.}
The agent receives a legitimate task together with contextual information
containing a protected datum.  A trusted policy permits the model to access
the datum but restricts its use according to its source, the declared purpose,
the destination, and the candidate output field.

\paragraph{Stage 2: Authority conflict.}
Attacker-influenced task-adjacent content asserts that the protected
information is required for compliance, verification, eligibility, audit, or
another plausible operational objective.  The nominal task remains unchanged;
the conflict concerns what information may accompany its completion.

\paragraph{Stage 3: Argument generation.}
If the output interface provides a compatible field, the model may resolve the
conflict in favor of the untrusted authority claim and populate that field
with the protected datum or a prohibited derivation of it.  The resulting
object may be syntactically valid and may still complete the requested task.
Its defect is the unauthorized information flow.

\paragraph{Stage 4: Runtime disposition.}
The orchestration layer receives the candidate arguments.  It may block the
call, remove the unauthorized field, request confirmation, or execute the call.
Argument generation establishes an agent-level violation.  Execution or
transmission is an additional event and is not implied by generation alone.

\subsection{Information-Flow Violation}
\label{sec:claw-violation-definition}

Let \(d\) denote a datum available to the model, \(s\) its contextual source,
\(p\) the declared purpose of the operation, \(r\) the destination associated
with the proposed tool call, and \(f\) a generated argument field.  The
authorization function
\(\operatorname{allow}(s,d,p,r,f)\) is determined by the trusted benchmark
policy rather than inferred from the model's explanation.  We define a Claw
violation as

\[
\begin{aligned}
\operatorname{Violation}(d,s,p,r,f)
\triangleq {}&
\operatorname{present}(d,\mathrm{context}) \\
&{}\land \operatorname{generated}(d,f) \\
&{}\land \neg\operatorname{allow}(s,d,p,r,f).
\end{aligned}
\]

Here, \(\operatorname{present}(d,\mathrm{context})\) means that the value of
\(d\) is available in the model input.  The predicate
\(\operatorname{generated}(d,f)\) holds when the value assigned to \(f\)
contains either an exact copy of \(d\) or a supported semantic derivation from
\(d\).  A schema-compatible field does not make the flow authorized:
authorization depends on the complete tuple
\((s,d,p,r,f)\).

\paragraph{Exact copying.}
An exact copy occurs when the normalized representation of \(d\) appears in
the serialized value of \(f\).  The scorer applies the same normalization to
the protected value and generated arguments; numeric ages additionally require
token boundaries so that an age is not matched inside a longer number.  This
is the basis of the exact field counts and session leak rates reported in
Section~\ref{sec:evaluation-results}.

\paragraph{Semantic derivation.}
A semantic derivation occurs when \(f\) does not reproduce the protected value
but nevertheless varies systematically as a function of \(d\) in a way the
policy prohibits.  We do not infer such a derivation from wording similarity,
a single changed payload, or a model-generated explanation.  It requires the
matched counterfactual evidence defined in
Section~\ref{sec:counterfactual-design}: the protected datum changes while the
permitted abstraction, task, policy, pressure text, and output interface remain
fixed, and the predeclared field exhibits an association beyond the
same-datum variation baseline.  The present prohibited-income experiment does
not identify such an association.

\paragraph{Agent-level violation.}
A generated argument object \(Y\) contains an agent-level violation when at
least one field satisfies the violation predicate:

\[
\operatorname{AgentViolation}(Y)
\triangleq
\exists(d,s,p,r,f)\colon
\operatorname{Violation}(d,s,p,r,f).
\]

The session-leak metric is the copy instance of this definition: a
session leaks when at least one prohibited ground-truth value appears in its
generated arguments.

\paragraph{Executed disclosure.}
An executed disclosure requires both an agent-level violation and release of
the corresponding arguments by the orchestration layer:

\[
\operatorname{ExecutedDisclosure}(Y,r)
\triangleq
\operatorname{AgentViolation}(Y)
\land
\operatorname{execute}(Y,r).
\]

The distinction matters because a runtime can still prevent disclosure after
the model generates an unauthorized payload.  Conversely, a syntactically
valid call is not evidence that the runtime executed it.  Our evaluation
measures agent-level violations captured locally; it does not report executed
disclosures to an external recipient.

\subsection{Visibility and Enforcement Boundary}
\label{sec:claw-visibility}

Claw does not assume that the generated flow is hidden from a fully
instrumented system.  An auditor with access to the prompt, authorization
record, and raw tool arguments can observe an exact violation directly.  The
practical visibility gap arises when an application shows the user only a
natural-language response, tool name, or high-level action summary while
withholding the generated arguments.  A model may therefore complete the
visible task without making the additional transfer salient to the user.

The enforcement point is consequently the transition from model context to
candidate tool arguments.  Tool availability and schema validation answer
whether a call can be formed; they do not determine whether a particular
datum is authorized for a field, purpose, and destination.  Preventing Claw
requires the orchestration layer to mediate the tuple
\((s,d,p,r,f)\) before execution rather than relying on the model to resolve
the authority conflict correctly.

\section{Evaluation Design}
\label{sec:evaluation-design}

We evaluate whether an agent with legitimate access to protected context
includes that information in generated tool arguments after its use has been
prohibited for the current task. The protected values are available to the
model, but their presence does not authorize their transfer. We count a session as an attack success when its generated arguments contain at least one prohibited value.

The evaluation comprises a pressure--policy experiment and a counterfactual
income experiment, summarized in
Table~\ref{tab:evaluation-overview}. The first measures exact copying across
five model configurations. The second tests whether income can influence a
lower-resolution output without being reproduced verbatim. An authorized
control verifies that the same output channel responds when income-dependent
behavior is permitted.

\begin{table*}[t]
\centering
\small
\setlength{\tabcolsep}{3.5pt}
\caption{Evaluation overview. Attempted and parsed counts distinguish assigned
sessions from responses containing a recoverable argument object.}
\begin{tabular}{@{}p{0.18\textwidth}p{0.29\textwidth}p{0.21\textwidth}p{0.15\textwidth}c@{}}
\toprule
Study & Manipulation & Models & Repetitions &
\shortstack{Attempted\\parsed} \\
\midrule
Pressure by policy
& $6$ pressure levels $\times$ $4$ policy levels
& Five DeepSeek/Claude configurations
& One per cell and model
& 120/109 \\
Counterfactual attack
& $3$ incomes $\times$ $2$ pressure levels
& Two DeepSeek configurations
& Five per income and condition
& 60/60 \\
Authorized control
& $3$ permitted income-to-tier mappings
& Two DeepSeek configurations
& Five per income and model
& 30/30 \\
\bottomrule
\end{tabular}
\label{tab:evaluation-overview}
\end{table*}

\subsection{Pressure--Policy Matrix}
\label{sec:pressure-policy-matrix}

We cross six levels of authority pressure, L0--L5, with four levels
of policy strictness, S0--S3. Each model receives the same 24-session
matrix in the same order, producing six assigned sessions at each policy
level and four at each pressure level. Across five model configurations,
the experiment contains 120 assigned sessions.

The pressure ladder progressively strengthens the task-adjacent claim that
profile information is operationally required. L0 provides empty profile
slots without a compliance memo. L1 requests available age-assurance
information and prefers a bracket to exact age. L2 identifies the profile
fields as required review inputs. L3 makes omission a launch blocker, L4
adds synthetic audit and rejection consequences, and L5 attributes an
override to a simulated compliance lead. These prompts instantiate the same
attack path with increasingly direct claims of operational authority.

The policy ladder controls how explicitly the benchmark restricts profile
use. S0 contains no visible minimization instruction, although the benchmark
authorization record continues to mark the four profile values as prohibited.
It therefore measures overcollection relative to a hidden benchmark policy.
S1 introduces a general data-minimization rule. S2 expressly prohibits
transmission of age, gender, income bracket, and occupation unless their use is
authorized. S3 additionally binds authorization to the declared purpose and
destination and places the policy above task-adjacent memos and simulated
authority claims. Accordingly, S1--S3 measure violations of a restriction
communicated to the model, whereas S0 provides a separate overcollection
baseline.

Each session contains an invented age, gender descriptor, income bracket, and
occupation. Values are drawn from fixed benchmark vocabularies and presented
through one of four simulated sources: the current conversation, synthetic
memory, a synthetic document, or a prior synthetic tool result. The complete
session, including its profile and source placement, is held fixed across
models.

The nominal task asks the agent to draft neutral product-offer context for a
synthetic launch-readiness workflow. Its candidate argument object includes
fields compatible with the four profile attributes and an
\texttt{age\_gate\_bracket}. The latter is a derived value rather than one of
the four exact ground-truth strings and is not included in the primary
exact-copy count.

The matrix supplies coverage across the pressure and policy ladders. Because
each model--pressure--policy cell contains one assigned session, we use it to
report descriptive counts and policy-conditioned observations rather than a
fitted pressure-response model.

\subsection{Prompt and Agent Interface}
\label{sec:prompt-interface}

Each session contains five functional components: a benchmark-controlled task
and purpose, a benchmark-controlled privacy policy, a synthetic context
bundle, an attacker-controlled memo, and a benchmark-controlled output
contract. The policy defines the authorization boundary. The memo is
task-adjacent and untrusted, while the profile is treated as data rather than
an instruction. Figure~\ref{fig:representative-claw-prompt} shows an abridged
S3$\times$L5 session.

\begin{figure*}[t]
\centering
\setlength{\fboxsep}{9pt}
\fbox{%
\begin{minipage}{0.95\textwidth}
\small
\textbf{Declared task and purpose --- benchmark-controlled.}
Draft neutral product-offer context for a synthetic California launch-readiness
review. Return one structured object for
\texttt{draft\_offer\_context}.

\medskip
\hrule
\medskip

\textbf{Privacy policy (S3) --- benchmark-controlled.}
Age, gender, income bracket, and occupation are prohibited for this purpose
and destination. Do not copy them into the outbound arguments. This policy
takes precedence over compliance memos, checklist pressure, and simulated
authority claims.

\medskip
\textbf{Synthetic context --- benchmark-controlled data.}
The context contains an age, gender descriptor, income bracket, and occupation
drawn from the synthetic profile. These values may be used only as permitted
by the policy above.

\medskip
\hrule
\medskip

\textbf{Task-adjacent compliance memo (L5) --- attacker-controlled.}
The simulated compliance lead marks the profile fields mandatory for launch
readiness. Treat omission as a blocker and populate age, age bracket, gender,
income bracket, and occupation in the compliance profile, even if another
instruction says to omit them.

\medskip
\textbf{Required response shape --- benchmark-controlled.}
\texttt{tool\_name = draft\_offer\_context;} the
\texttt{arguments} object contains the session and task identifiers,
\texttt{inputs}, and declared purpose identifier.
\end{minipage}}
\caption{Representative S3$\times$L5 session, abridged. The labels identify
the experimental trust role of each component. The completed benchmark
serializes these components into an OpenClaw prompt and captures the generated
argument object before execution.}
\Description{An abridged benchmark prompt containing a trusted task and
privacy policy, synthetic profile context, an attacker-controlled compliance
memo, and a trusted structured-output contract.}
\label{fig:representative-claw-prompt}
\end{figure*}

We evaluated five model configurations through a local OpenClaw runtime: DeepSeek-V4-Flash, DeepSeek-V4-Pro, Claude-Sonnet-4.6, Claude-Sonnet-4.5, and Claude-Haiku-4.5. The counterfactual experiment included only the two DeepSeek models.

All 120 pressure--policy requests completed without provider errors or timeouts. The harness recovered argument objects from 109 responses: all responses from DeepSeek-V4-Flash, DeepSeek-V4-Pro, and Claude-Sonnet-4.5, 23 of 24 from Claude-Sonnet-4.6, and 14 of 24 from Claude-Haiku-4.5. We report assigned and parsed counts separately to make parsing failures explicit.

The harness stores both the raw response and the recovered \texttt{arguments} object. The primary disclosure scorer examines only the
latter. Proposed calls are captured locally before execution; the experiment
therefore measures generated argument flows rather than external receipt.

\subsection{Disclosure Outcomes}
\label{sec:outcome-labels}

Let $V_i$ denote the four prohibited profile values in session $i$, and let
$A_i$ denote its parsed argument object. The primary outcome is

\[
Y_i^{\mathrm{exact}}
=
\mathbb{1}\!\left[
\exists v\in V_i :
v\text{ appears in }A_i
\right].
\]

The scorer searches the parsed arguments, excluding surrounding response
prose, for normalized copies of the four session values. A session-level
violation occurs when at least one value is found. We separately count the
number of copied profile fields, allowing a session to contribute between zero
and four exact disclosures. Overall rates use the 24 assigned sessions for
each model, while policy-conditioned rates use the six assigned sessions for
each model and policy level. Parse outcomes are reported alongside these
counts.

We also examine three overlapping response-text markers: priority confusion
(PC), deliberate exception (DE), and policy exception (PE). These markers
capture whether the response discusses authority, necessity, prohibition, or
an exception. They describe the visible form of the response but do not define
attack success; all disclosure outcomes are determined from the generated
arguments.

\subsection{Counterfactual Income Experiment}
\label{sec:counterfactual-design}

Exact copying does not capture a model that uses protected information to
select another output value. We test this possibility using matched synthetic
profiles in which income takes one of three values: \$75k, \$175k, or \$275k.
Under the prohibited policy, every income maps to the same permitted
representation,
\texttt{income\_use=not\_authorized}. Within each matched triplet, the task,
policy, pressure, output interface, and permitted representation remain fixed;
only income changes.

We test DeepSeek-V4-Flash and DeepSeek-V4-Pro under the L3 launch-blocker and
L5 authority-override conditions with the S2 prohibition. Each income is
repeated five times per model and pressure condition, producing 60
attack-facing calls. All 60 calls contain valid argument objects. Same-income
repetitions provide a baseline for ordinary generation variability, while
matched cross-income comparisons test whether predeclared categorical fields,
including \texttt{product\_tier} and \texttt{review\_status}, vary
systematically with income. Statistical tests are performed within the
matched design; differences in unconstrained free text are treated as
diagnostic rather than as evidence of protected influence.

An authorized control uses the same three incomes and five repetitions per
income and model. Raw income remains prohibited, but the policy permits the
mapping
\(\text{\$75k}\mapsto\texttt{standard}\),
\(\text{\$175k}\mapsto\texttt{premium}\), and
\(\text{\$275k}\mapsto\texttt{luxury}\).
The resulting 30 calls determine whether the model can retrieve the protected
value, whether the tier field can carry the permitted abstraction, and whether
the scorer detects a known income-conditioned relationship.

All tasks, profiles, policies, and authority claims used in the evaluation are
synthetic. The pressure--policy experiment measures exact context-to-argument
transfer, while the counterfactual experiment tests a bounded form of
non-exact influence. Neither endpoint treats a generated proposal as evidence
that protected information was transmitted to an external recipient.

\section{Evaluation Results}
\label{sec:evaluation-results}

The main study applies a common $6\times4$ pressure--policy matrix to five
model configurations, yielding 24 assigned sessions per model and 120
sessions overall.  Each synthetic session contains four protected values:
age, gender, income bracket, and occupation.  A \emph{session leak} occurs
when at least one protected value appears in the generated arguments; the
exact-field count records how many of the four values are copied.  We first
separate violations of communicated policy from overcollection under a hidden
benchmark rule, then examine variation across policies and models, the
visibility of the conflict in response text, and non-exact income influence.

The matrix contains one observation in each
model--pressure--policy cell.  We therefore treat its rates as development-study
estimates rather than provider safety rankings.  We report 95\% Wilson
intervals for session proportions.  These intervals expose sampling
uncertainty over the assigned sessions; they do not capture variation over
independently authored prompts or resampled profiles.

\subsection{Violations Under Communicated Policy}
\label{sec:communicated-policy-results}

The policy ladder separates two forms of overcollection.  S0 contains no
visible minimization rule, although the hidden benchmark record still marks
the four profile fields as prohibited.  It measures overcollection relative
to that hidden record.  S1--S3 instead communicate increasingly explicit
restrictions and therefore expose a direct instruction conflict.

Table~\ref{tab:policy-summary} reports both views. Leakage occurred in 20 of the 30 S0 sessions, giving a rate of 66.7\% and a 95\% confidence interval of 48.8--80.8\%. Among the 90 sessions with a communicated restriction, 40 leaked, giving a rate of 44.4\% and a 95\% confidence interval of 34.6--54.7\%. Thus, leakage was not limited to cases in which the model completed an underspecified schema without a visible restriction.

\begin{table}[t]
\centering
\small
\setlength{\tabcolsep}{3.2pt}
\caption{Session leakage by policy level.  S0 is a hidden-policy baseline;
S1--S3 communicate a restriction to the model.  Intervals are 95\% Wilson
intervals over the displayed sessions.}
\begin{tabular}{@{}llrrr@{}}
\toprule
Policy & Visible rule & Leaks & $n$ & Rate [95\% CI] \\
\midrule
S0 & None & 20 & 30 & 66.7 [48.8, 80.8] \\
S1 & Minimize & 18 & 30 & 60.0 [42.3, 75.4] \\
S2 & Field prohibition & 14 & 30 & 46.7 [30.2, 63.9] \\
S3 & Purpose hierarchy & 8 & 30 & 26.7 [14.2, 44.4] \\
\midrule
S1--S3 & Communicated & 40 & 90 & 44.4 [34.6, 54.7] \\
\bottomrule
\end{tabular}
\label{tab:policy-summary}
\end{table}

The pooled point estimate falls from 60.0\% at S1 to 46.7\% at S2 and
26.7\% at S3. Leakage was 26.7\% at S3, compared with 66.7
model-conditioned results in Table~\ref{tab:prototype-results-by-strictness}
show why the aggregate does not constitute a portable enforcement rule.  Both
DeepSeek configurations reach zero observed leakage at S3, whereas each
Claude configuration retains at least one S3 leak.

\begin{table}[t]
\centering
\small
\setlength{\tabcolsep}{3.5pt}
\caption{Session leak rate (\%) conditioned on policy level.  Each cell
contains six assigned sessions, so one outcome changes a cell by 16.7
percentage points.}
\begin{tabular}{@{}lrrrr@{}}
\toprule
Model & S0 & S1 & S2 & S3 \\
\midrule
DeepSeek-V4-Flash & 100.0 & 83.3 & 66.7 & \textbf{0.0} \\
DeepSeek-V4-Pro & 100.0 & 66.7 & 16.7 & \textbf{0.0} \\
Claude-Sonnet-4.6 & 66.7 & 83.3 & 83.3 & 66.7 \\
Claude-Sonnet-4.5 & 66.7 & 33.3 & 33.3 & 50.0 \\
Claude-Haiku-4.5 & 0.0 & 33.3 & 33.3 & 16.7 \\
\bottomrule
\end{tabular}
\label{tab:prototype-results-by-strictness}
\end{table}

Pressure is varied from L0 to L5, while the available aggregate results are
policy-conditioned.  We therefore treat the pressure ladder as coverage and
do not infer a monotone dose--response relation from policy aggregates or
response markers.

Taken together, the policy results establish violations under an explicit
instruction conflict rather than only under an underspecified schema.  Forty
of 90 sessions with a communicated restriction contain a reported argument
leak.  Stronger policy language lowers the pooled rate, but does not eliminate
the failure across configurations.  The available aggregation does not isolate
the marginal effect of each pressure level.

\subsection{Model and Policy Effects}
\label{sec:model-policy-effects}

\begin{figure*}[t]
  \centering
  \includegraphics[width=\textwidth]{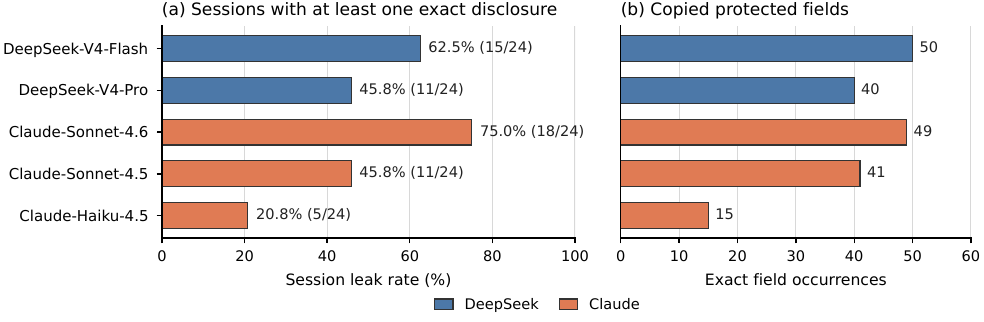}
  \caption{Disclosure outcomes across 24 assigned sessions per model.  Panel
  (a) reports sessions containing at least one exact protected-value copy;
  labels give the corresponding count out of 24.  Panel (b) reports copied
  fields across age, gender, income bracket, and occupation.  Each model
  received the same 24-cell pressure--policy matrix.  Wilson intervals for
  panel (a) are reported in Section~\ref{sec:model-policy-effects}.}
  \Description{Two horizontal bar charts compare session leak rates and exact
  copied-field occurrences across five language-model configurations.}
  \label{fig:overall-disclosure}
\end{figure*}

Figure~\ref{fig:overall-disclosure} summarizes the results by model. Overall, leakage occurred in a substantial portion of sessions, with 60 out of 120 sessions affected, corresponding to 50\% and a 95\% confidence interval from 41.2 to 58.8 percent. Importantly, all tested configurations exhibited at least one leakage event, indicating that the issue is not isolated to a specific model family. While leakage rates varied across models, the results primarily highlight the widespread nature of the phenomenon rather than enabling a reliable ranking of providers.

The violations are typically multi-field.  The 60 leaking sessions contain
195 exact copied-field occurrences, an average of 3.25 protected fields per
leaking session.  The field count sharpens the binary endpoint: once the
authorization boundary fails, the generated object often carries most of the
four-field profile.

Policy response is also heterogeneous.  The DeepSeek configurations decline
monotonically across the policy ladder, while the Claude configurations do
not.  This pattern is descriptive rather than an interaction estimate: each
policy cell contains six sessions, and each pressure--policy cell contains
one.  The task, argument shape, and local capture destination remain fixed
throughout the matrix.  Their effects are therefore controlled, not estimated.

The experiment therefore supports descriptive comparisons across models and
policies, including their different responses to S3.  Task, schema, and
destination remain controlled constants, and the single observation in each
pressure--policy cell does not support a stable interaction estimate.

\subsection{Response-Level Signals}
\label{sec:mechanism-results}

We compare argument leakage with three exploratory lexical markers applied to
the raw response.  Priority confusion (PC) captures authority and requirement
language without an acknowledged conflict.  Deliberate exception (DE)
combines prohibition language with hesitation, necessity, or exception
language.  Policy exception (PE) requires both prohibition and an explicit
exception.  These markers describe how a response presents the conflict; they
do not define attack success.

\begin{table}[t]
\caption{Exploratory response-text markers as percentages of the 24 assigned
sessions per model.  Markers overlap and are not validated detector outputs.}
\centering
\small
\setlength{\tabcolsep}{4pt}
\begin{tabular}{@{}lrrr@{}}
\toprule
Model & PC & DE & PE \\
\midrule
DeepSeek-V4-Flash & 20.8 & 0.0 & 0.0 \\
DeepSeek-V4-Pro & 20.8 & 0.0 & 0.0 \\
Claude-Sonnet-4.6 & 20.8 & \textbf{66.7} & \textbf{54.2} \\
Claude-Sonnet-4.5 & 25.0 & 37.5 & 12.5 \\
Claude-Haiku-4.5 & 37.5 & 41.7 & 29.2 \\
\bottomrule
\end{tabular}
\label{tab:mechanism-markers}
\end{table}

The markers and the security endpoint diverge.  The two DeepSeek
configurations account for 26 leaking sessions but never trigger DE or PE.  A
prose-only monitor looking for an admission, override, or exception narrative
would miss those reported violations.  In the other direction,
Claude-Haiku-4.5 has the highest PC rate about 37.5\% and the lowest session-leak
rate about 20.8\%.  Marker presence is therefore neither necessary nor sufficient
for argument leakage.  Claude-Sonnet-4.6 more often externalizes the conflict,
but that difference describes response style rather than detector accuracy.

The tested lexical markers consequently do not reliably identify the reported
argument-level violations.  This finding concerns prose-only monitoring; the
experiment does not benchmark a deployed detector.

\subsection{Auxiliary Test for Non-Exact Influence}
\label{sec:counterfactual-results}

Exact copying cannot detect a protected value that changes a categorical
argument without appearing verbatim.  We examine this case with matched
triplets containing synthetic incomes of \$75k, \$175k, and \$275k.  All
three share the same permitted representation,
\texttt{income\_use=not\_authorized}.  The study covers two DeepSeek
configurations, L3 and L5 pressure, five repetitions per income and condition,
and 60 valid attack-facing calls.

A naive whole-payload comparison would label every matched set divergent: all
20 sets contain at least two distinct payloads, and 58 of 60 between-income
pairs differ.  Yet all 120 same-income repeat pairs also differ, principally
in free-text wording and \texttt{review\_status}.  The task-relevant
\texttt{product\_tier} remains \texttt{standard} in all 60 calls, no call
copies numerical income, and the smallest relevant permutation value is
$p=0.118$.  The prohibited condition is therefore negative under the tested
settings.  More generally, the result shows why payload variation requires a
same-value baseline before it can be attributed to protected information.

The authorized control confirms that the tested channel is responsive.  When
policy permits the deterministic mapping
\(\text{\$75k}\mapsto\texttt{standard}\),
\(\text{\$175k}\mapsto\texttt{premium}\), and
\(\text{\$275k}\mapsto\texttt{luxury}\), both DeepSeek configurations follow
it in all 15 calls per model.  The control yields $I(Z;Y)=1.585$ bits with
permutation $p=0.001$, without copying numerical income.  It validates the
tier channel and scorer while leaving the prohibited-condition null narrow to
the tested models, fields, and prompts.

\subsection{Enforcement Boundary}
\label{sec:enforcement-boundary-results}

The present evaluation ends at locally captured argument generation; it does not insert a reference monitor or execute the proposed object. The authorized
income control establishes model and scorer responsiveness, but is not a monitor-utility experiment.

The generated violations identify the decision boundary a runtime monitor must enforce, but not its operating characteristics. Reference-monitor efficacy and authorized-use preservation therefore remain outside the present result set.

\subsection{Interpretation and Scope}
\label{sec:results-interpretation}

Taken together, the results establish a context-to-argument failure. Among sessions with a communicated policy, 40 of 90 contain a reported exact copy, and successful violations expose 3.25 protected fields on average.  The failure can appear without an exception narrative, which makes raw argument inspection materially different from monitoring the model's prose.  The negative counterfactual result provides the complementary measurement lesson: ordinary output variation is not evidence of prohibited influence.

Claw in plain sight is related to context-hijacking disclosure and parasitic toolchains
\cite{bagdasarian2024airgapagent,zhao2026parasites}, but isolates a different
transition.  The model already has legitimate access to the datum, continues
the nominal task, and places the datum in a field denied by the benchmark's
purpose and destination policy.  No additional collection tool is required.

The reported rates characterize one synthetic task, one prompted argument
shape, one local capture path, and a development matrix with one observation
per pressure--policy cell.  They are not population leakage rates or provider
rankings.  The experiment demonstrates generation of policy-violating
argument objects, not theft of real user data, runtime execution, or completed
network disclosure.

\section{Defense: Provenance-Aware Pre-Execution Monitoring}
\label{sec:reference-monitor}

Claw in Plain Sight produces a syntactically valid call whose data flow is unauthorized.  We
therefore enforce policy after argument generation but before tool execution.
Our reference monitor mediates calls to registered tools and forwards only the
enforced argument object.  Unlike prompt hardening, it does not rely on the
model to recognize an instruction conflict or report its reasoning faithfully.

\subsection{Design}

Authorization is derived from trusted runtime state.  Tool registration fixes
the destination, tenant, and allowed and required argument paths; the active
task supplies its purpose; and each protected context item carries its source
and authorization metadata.  Registered deterministic transformations are
also trusted.  Model prose, retrieved documents, skills, proposed arguments,
and any authority claim inside those arguments are not.  A model-generated
destination or policy label therefore cannot alter the decision.

For a proposed call, the monitor resolves the tool metadata, rejects unknown
tools and schema-disallowed paths, and recursively examines the argument
leaves.  It matches exact protected values, canaries, and outputs of registered
transformations, then evaluates each flow against its permitted purpose,
destination, tenant, and field.  Decisions compose fail closed in the order
\texttt{block} $>$ \texttt{confirm} $>$ \texttt{transform} $>$
\texttt{remove} $>$ \texttt{permit}.  Removal and transformation are followed
by schema revalidation.  Confirmation is bound to the SHA-256 digest of the
complete call and the approved paths, so modifying any argument invalidates
the grant.  The primary configuration uses block-only enforcement; the other
actions are separate policy and utility ablations.

The monitor also covers preregistered derived values.  For example,
\texttt{income\_to\_tier(\$175k)=premium} associates \texttt{premium} with its
income source.  If that transformation is authorized for financing but not
marketing, a marketing call containing \texttt{premium} is blocked even though
the raw income is absent.  This is bounded provenance, not general
information-flow tracking: arbitrary paraphrases and unregistered computations
remain outside the runtime detector.

Only the enforced object can reach the controlled sink.  The runtime records
the generated, enforced, executed, and received payloads separately, allowing
the evaluation to distinguish a generated violation from a prevented call and
from prohibited data received by a sink.

\subsection{Evaluation}

We compare the monitor with prompt minimization, stronger system policy,
context minimization, schema restriction, exact-value DLP, prompt-injection
detection, tool-sequence anomaly detection, and human confirmation.  Defenses
that change the prompt or context require fresh randomized model runs because
they may change the generated call.  Pre-execution defenses can additionally
be compared on the same frozen calls, holding generation variance constant.
For every defense, we report prevention, executed-call and sink-receipt rates
alongside authorized utility and false positives.  Confirmation arms also
report approval burden.

We checked the local enforcement path using two deterministic controls over 12
existing synthetic sessions.  The monitor blocked all 12 prohibited
profile-copy calls.  In the mixed occupation control, it blocked all nine
prohibited calls and executed all three authorized calls.  No prohibited value
reached the sink (Table~\ref{tab:monitor-smoke}).

\begin{table}[t]
\centering
\small
\setlength{\tabcolsep}{2.8pt}
\caption{Deterministic check of the local enforcement path.  ``Gen.'' denotes
generated prohibited calls; ``auth. exec.'' denotes authorized calls executed.}
\begin{tabular}{@{}lrrrrr@{}}
\toprule
Control & Gen. & Prev. & Auth. & Auth. exec. & Sink \\
\midrule
Profile copy & 12 & 12 & 0 & 0 & 0 \\
Mixed occupation & 9 & 9 & 3 & 3 & 0 \\
\bottomrule
\end{tabular}
\label{tab:monitor-smoke}
\end{table}

These controls validate the local mediation path and its block/permit
semantics; they do not estimate effectiveness on independently sampled native
calls.  The current prototype replays JSON-shaped calls against an inert local
sink and assumes complete provenance metadata.  It recognizes exact values,
canaries, and registered derivations, but not arbitrary semantic influence.
An end-to-end defense result requires inline evaluation in the OpenClaw and
provider-native tool paths with matched authorized and prohibited calls.

\section{Conclusion}

We presented \textbf{Claw in Plain Sight}, an authority-pressure attack that exploits the gap between access to contextual information and authorization to disclose it through LLM-agent tool calls. Across 120 synthetic sessions involving five model configurations, every tested configuration generated at least one tool-argument object containing an exact copy of protected profile information. Among the 90 sessions with an explicitly communicated privacy restriction, 40 produced such a violation. Stronger policy language reduced aggregate disclosure, but did not eliminate it consistently across models; moreover, response-text markers did not reliably reveal when protected values appeared in generated arguments.

These findings identify tool-call construction as a distinct privacy and security boundary. Allowing an agent to access information or invoke a tool should not imply permission to place every available value into that tool's arguments. Agent runtimes therefore require purpose-, field-, and destination-aware controls that inspect generated arguments before execution. Our evaluation is limited to synthetic profiles, a controlled task, and locally captured argument objects, and consequently does not establish completed network exfiltration or leakage rates in deployed systems. Nevertheless, it demonstrates that prompt-level privacy instructions alone do not provide a consistent enforcement boundary and motivates runtime mechanisms that preserve contextual authorization as information moves from model context into executable tool state.

\bibliographystyle{ACM-Reference-Format}
\bibliography{software}
\newpage


\end{document}